\documentstyle[12pt]{article}
\def\be{\begin{equation}}
\def\ee{\end{equation}}
\def\bea{\begin{eqnarray}}
\def\eea{\end{eqnarray}}
\def\beqa{\begin{eqnarray}}
\def\eeqa{\end{eqnarray}}
\def\beq{\begin{equation}}
\def\eeq{\end{equation}}
\def\beqal{\begin{eqnarray}\label}
\def\beql{\begin{equation}\label}

\def\R{\mbox{\rm I\kern-.18em R}}

\def\P{\mbox{\rm I\kern-.18em P}}
\def\uno{\mbox{1 \kern-.59em {\rm l}}}
\def\Ds{\ {\big / \kern-.70em D}}
\def\ds{\big / \kern-.90em {\ \p} }

\def\cDs{\ {\big / \kern-.70em {\cal D}}}
\def\Z{{Z \kern-.45em Z}}
\def\Q{{\kern .1em {\raise .47ex \hbox{$\scriptscriptstyle |$}}
\kern -.35em {\rm Q}}}

\def\Tr{\mbox{\rm Tr}}

\def\p{\partial}

\def\om{\omega}

\def\1{\dot{1}}
\def\2{\dot{2}}

\font\mybb=msbm10 at 12pt

\def\bb#1{\hbox{\mybb#1}}
\def\Z {\bb{Z}}

\begin{document}\begin{titlepage}

\hfill{DFPD01/TH/13}

\hfill{hep-th/0105041}

\vspace{1cm}
\begin{center}
{\large \bf THE INSTANTON UNIVERSAL MODULI SPACE OF}

\vspace{.3cm}

{\large\bf N=2 SUPERSYMMETRIC YANG-MILLS THEORY}

\end{center}
\vspace{1.5cm}
\centerline{MARCO MATONE}
\vspace{0.8cm}
\centerline{\it Department of Physics ``G. Galilei'' - Istituto Nazionale di
Fisica Nucleare}
\centerline{\it University of Padova}
\centerline{\it Via Marzolo, 8 - 35131 Padova, Italy}
\centerline{matone@pd.infn.it}

\vspace{2cm}
\centerline{\sc ABSTRACT}

\vspace{0.6cm} \noindent We use the recursive structure of the
compactification of the instanton moduli space of $N=2$ Super
Yang-Mills theory with gauge group $SU(2)$ to construct, by an
inductive limit, a universal moduli space which includes all the
multi-instanton moduli spaces. Furthermore, by exploiting an analogy 
with the strong and weak coupling expansions in the Matrix Model 
formulation of 2D Quantum Gravity, we discuss the possible field 
theoretical meaning of the contributions to the prepotential in the 
strong coupling region. In particular, whereas the weak and strong 
coupling expansions of the Painlev\'e I correspond to the genus and 
punctured spheres expansions, in the case of the instantons there 
should exist a space dual to the instanton moduli space.

\end{titlepage}
\newpage
\setcounter{footnote}{0}
\renewcommand{\thefootnote}{\arabic{footnote}}
\noindent Recently in \cite{M1} it has been shown that the
instanton moduli spaces of $N=2$ Super Yang-Mills theory with
gauge group $SU(2)$ admit a compactification which is at the heart
of the recursive relations derived in \cite{alfa}. The structure
of this compactification explains why, in spite of the technical
difficulties, the Seiberg-Witten (SW) solution \cite{sw} is
simple. In the SW model there exists a relation between the
modulus $u=\langle\Tr\,\phi^2\rangle$ and the effective
prepotential \cite{alfa} (see also \cite{vari}), which allowed to
prove the SW conjecture by using the reflection symmetry of
quantum vacua \cite{BMT}. Thus, essentially, the research started
{}from the exact solution and ended with the reconstruction of the
field theoretic structure of the theory.

\noindent A basic point in \cite{M1} was the observation of the
strict similarity between the properties of the instanton moduli
space and its measure on the one hand and the corresponding
quantities in the theory of punctured spheres, on the other. In
particular, it has been shown that a compactification similar to the 
one by Deligne-Knudsen-Mumford \cite{DeligneKnudsenMumford} and the 
restriction phenomenon satisfied by the Weil-Petersson volume form 
\cite{wolpertis} lead to the recursive structure obtained in the SW 
model. The coefficients which determine the divisors at the boundary 
of the instanton moduli space and the structure of the instanton 
2-form are fixed by the coefficients defining the recursion 
relations \cite{alfa}. This mechanism is similar to the one 
considered in \cite{ma,bomama}. In this respect the appearance of 
linear differential equations plays a crucial role 
\cite{KLT,M1,alfa,manin}. According to \cite{manin}, this may be 
related to a sort of mirror phenomenon. In particular, we recall 
that the recursion relations satisfied by the instanton 
contributions and by the Weil-Petersson volume forms are fixed by an 
underlying linear differential equation. In \cite{M1} it has been 
shown that the instanton contribution to \beq
\langle\Tr\,\phi^2\rangle=a^2\sum_{n=0}^\infty {\cal G}_n
\left({\Lambda\over a}\right)^{4n}\ \ , \label{ux1}\eeq can be
expressed as \beq\label{uno} {\cal 
G}_{n}=\int_{\overline{V}_I^{(n)}}\bigwedge_{k=1}^{2n-1}
\om_I^{(n)}\ \ . \label{cancheck}\eeq 
The 2-form $\om_I^{(n)}$ is defined on $\overline{V}_I^{(n)}$ 
denoting a suitable compactification of the $n$-instanton moduli 
space ${V}_I^{(n)}$. This result follows from some algebraic 
geometrical calculations whose details are given in \cite{M1}. Let 
us stress that the integral reproduces the recursion relations 
\cite{alfa,M1} \beq {\cal G}_{n+1}={1\over 2(n+1)^2} 
\left[(2n-1)(4n-1){\cal G}_n +\sum_{k=0}^{n-1}b_{k,n}{\cal 
G}_{n-k}{\cal G}_{k+1} 
-2\sum_{j=1}^{n-1}\sum_{k=1}^{j}d_{j,k,n}{\cal G}_{n-j} {\cal 
G}_{j+1-k}{\cal G}_{k}\right]\ \ , \label{recursion3} \eeq where
$n>0$, ${\cal G}_1=1/4$,
$b_{k,n}=c_{k,n}-2d_{k,0,n}$ and \beq c_{k,n}=2k(n-k-1)+n-1\
\ , \qquad d_{j,k,n}= [2(n-j)-1][2n-3j-1+2k(j-k+1)]\ \ . \eeq
The above result is due to two basic properties. First, as we said, 
the compactification of ${V}_I^{(n)}$ is similar to the 
Deligne-Knudsen-Mumford compactification. Actually, this has an 
obvious recursive structure. In the case of instantons we have that 
the boundary of the $(2n-1)$-dimensional space $\overline V^{(n)}_I$ 
decomposes as \cite{M1} \beq {\cal 
D}^{(n+1)}=\overline{V}_I^{(n+1)}/V_I^{(n+1)}=
\sum_{j=0}^{n-1}{\cal D}_{1,j}+\sum_{j=1}^{n-1}\sum_{k=1}^{j}
{\cal D}_{2,j,k}+{\cal D}_{3,n} \ \  , \label{bhos}\eeq where
\beqa {\cal D}_{1,j}&=&c^{(1)}_{n,j}\overline{V}_I^{(n-j)}\times
\overline{V}_I^{(j+1)}\ \ ,\nonumber\\
{\cal D}_{2,j,k}&=&c^{(2)}_{n,j,k}
\overline{V}_I^{(n-j)}\times\overline{V}_I^{(j+1-k)}\times
\overline{V}_I^{(k)}\times
\overline{V}_I^{(1)}
\ \ ,\nonumber\\
{\cal D}_{3,n}&=&c^{(3)}_{n}\overline{V}_I^{(n)}\times \overline{ 
V}_I^{(1)}\ \ . \label{ubisse}\eeqa Note that ${\cal D}_{3,n}$ can 
be included either in ${\cal D}_{1,0}$ or ${\cal D}_{1,n-1}$ by 
changing the coefficients. The second relevant property we used to 
reproduce the recursion relations is the restriction phenomenon. 
This is a property satisfied by the Weyl-Petersson 2-form. We 
defined $\om_I^{(n)}$ in such a way that a similar phenomenon occurs 
also in the present framework. Roughly speaking, this phenomenon 
consists in the property that the restriction of $\om_I^{(n)}$ to a 
component, e.g. to $\overline{V}_I^{(k)}$, of the boundary of 
$\overline{V}_I^{(n)}$, is in the same cohomological class of
$\om_I^{(k)}$. By simple calculations it is then easy to see
that these two properties allow to reproduce the recursion 
relations. The parameters defining $\om_I^{(n)}$ and
$\overline{V}_I^{(n)}$ are fixed by the parameters 
of the recursion relations themselves. Let us then recall how 
$\om_I^{(n)}$ is constructed. Let ${\cal D}^{(n+1)}_{\om}$ be the 
$4n$-cycle corresponding to the Poincar\'e dual to the ``instanton'' 
class $[\om_I^{(n+1)}]=c_1([{\cal D}^{(n+1)}_{\om}])$, where $[{\cal 
D}]$ is the line bundle associated to a given divisor ${\cal D}$ and 
$c_1$ denotes the first Chern class. In terms of the divisors at the 
boundary of the moduli space we have \beq\label{boh} {\cal 
D}^{(n+1)}_{\om}=\sum_{j=0}^{n-1}d^{(1)}_{n,j}{\cal D}_{1,j}+
\sum_{j=1}^{n-1}\sum_{k=1}^{j}d^{(2)}_{n,j,k}{\cal D}_{2,j,k}+
d^{(3)}_{n}{\cal D}_{3,n}\ \ . \eeq The coefficients satisfy the
relations \cite{M1} \beqa
&&d^{(1)}_{n,k}c^{(1)}_{n,k}\pmatrix{2n\cr 2(n-k)-1}=
\frac{b_{k,n}}{2(n+1)^2}
\ \ ,\nonumber\\
&&d^{(2)}_{n,j,k}c^{(2)}_{n,j,k}\pmatrix{2n\cr 2k}
\pmatrix{2(n-k)\cr 2(n-j)-1}=-\frac{2d_{j,k,n}}{k(n+1)^2}\ \ ,
\nonumber\\
&&d^{(3)}_{n}c^{(3)}_{n}=\frac{(2n-1)(4n-1)}{n(n+1)^2}\ \ . \eeqa

\noindent An open problem in SW theory consists in understanding
the field theoretical characterization of the theory in the strong
coupling region. In particular, it is not clear how it is possible
to determine, by purely field theoretical means, the coefficients in 
the expansion of the prepotential near $u=\pm\Lambda^2$. Thus, it 
would be desirable to find expressions representing, in the strong 
coupling region, the analogues of the integrals on the instanton 
moduli spaces. Usually, the theory in the strong coupling region is 
investigated by considering the corresponding dual Abelian theory. 
However, the meaning of the coefficients in the expansion of 
${\cal F}_D$ is still unclear. On the other hand, from a mathematical
point of view, the expansion of ${\cal F}$ itself in the strong 
coupling region is well defined. In this context, we observe that 
the SW solution is essentially fixed by the transformation 
properties of the prepotential under duality. In other words, the 
transformation properties of the prepotential essentially fixes 
${\cal F}$ itself. In turn, it determines all the instanton 
contributions. A possible way to investigate the dual space is to 
relate the two regions by deforming the instanton moduli space. In 
this way we should obtain the corresponding space near to 
$u=\pm\Lambda^2$. A natural approach is to consider
the Borel sum. In particular, in the following we will express 
$u=\langle\Tr\,\phi^2\rangle$ as an integral on an infinite 
dimensional space which depends on a parameter on which we will 
perform the Borel sum. This is a first step towards expressing 
relevant quantities on spaces which, outside the weak coupling 
region, do not correspond to instanton moduli spaces. A similar 
expression for the prepotential ${\cal F}$ can be obtained from 
$u$ by using \cite{alfa} \beq u=\pi i \left({\cal F}-a^2{\partial
{\cal F}\over\partial a^2}\right)\ \ ,
\label{arelazione}\eeq which is equivalent to \beq {\cal
F}(a)=a^2\left({2i\over \pi}\int^a_{a_0} dy y^{-3}{\cal
G}(y)+{a_{D0}\over 2a_0}-{i\over \pi}{u_0\over a^2_0}\right)\ \ ,
\label{ffunzofu}\eeq where $u=\Lambda^2{\cal G}(a)$, while $u_0$
and $a_{D0}$ correspond to $u$ and $a_D$ at $a_0$.

\noindent Although we know the differential equation satisfied by
$\langle\Tr\,\phi^2\rangle$ and ${\cal F}$, as we said,
we are interested in deriving their expression outside the weak 
coupling region by assuming Borel summability. Thus, we will 
consider the expression \beq 
\langle\Tr\,\phi^2\rangle=a^2\int_0^\infty dx e^{-x} 
\sum_{n=0}^\infty {1\over n!}{\cal G}_n x^n\left({\Lambda\over 
a}\right)^{4n}\ \ , \label{uy1}\eeq where
${\cal G}_0=1/2$. The precise domain of this Borel sum is related
to an asymptotic analysis which we will consider elsewhere. One
may also investigate other analytic continuations.

\noindent We now define the ``instanton universal moduli space''
$\overline{\cal V}^{(\infty)}(q)$. Let us consider the embedding
\begin{equation}
i_n:\overline V^{(n)}_I\longrightarrow \overline V^{(n+1)}_I,\qquad
n>1\ \ .
\label{u1}\end{equation}
Let $q\in {\R}_+$, we define by inductive limit
\begin{equation}
\overline{\cal V}^{(\infty)}(q)=\coprod_{n=0}^\infty\left(
\overline V^{(n)}_I\times [0,q^n]\right)/ \left(\overline
V^{(n)}_I,q^n\right)\sim \left(i_n \overline V^{(n)}_I,0\right)\ \ , 
\label{u2}\end{equation} where $\overline V^{(0)}_I$ is a point. Let 
$dy$ denote the Lebesgue measure on the real axis and consider the 
indefinite rank forms
\begin{equation}
\eta^{(\infty)}=\left[{1\over
2}+\sum_{n=1}^\infty{\omega^{(n)^{2n-1}}_I\over n!} \right]\wedge
dy\ \ . \label{u3}\end{equation} We then have
\begin{equation}
\langle {\rm Tr}\, \phi^2 \rangle=a^2\int_0^{\infty} dx
e^{-x}\int_{\overline{\cal V}^{(\infty)}\left(x{\Lambda^4/
a^4}\right)} \eta^{(\infty)}\ \ . \label{u5}\end{equation} This
expression should correspond to the one obtained by purely field
theoretical means. This would provide a possible approach to
understanding the role of the higher order contributions to ${\cal 
F}_D$. In this respect we observe that the moduli space in 
(\ref{u5}) depends on $x$. We note that, in principle, similar 
moduli spaces in the strong coupling region can be constructed by 
using the recursion relations which follow by expanding the 
prepotential at $u=\pm\Lambda^2$.

\noindent In the case in which $a\to\infty$, Eq.(\ref{u5})
reproduces the usual result in the form of an integral on the
infinite-dimensional universal moduli space. In particular, we
have
\begin{equation}
\langle {\rm Tr}\, \phi^2 \rangle=a^2\int_{\overline{\cal
V}^{(\infty)}\left(x{\Lambda^4/a^4}\right)} \omega^{(\infty)}\ \ ,
\label{u6}\end{equation} where
\begin{equation}
\omega^{(\infty)}=\left[{1\over
2}+\sum_{n=1}^\infty\omega^{(n)^{2n-1}}_I \right]\wedge dy\ \ .
\label{u4}\end{equation} We observe that the appearance of a
recursive structure similar to the one considered in the
Deligne-Knudsen-Mumford compactification suggests a deep relation
between the geometry of 4D SYM theory and the theory of Riemann
surfaces. On the other hand, we already encountered a similar
structure when the role of uniformization (Liouville) theory
appeared to be relevant in the SW solution \cite{alfa,BMT}.

\noindent The similarity between this approach and the theory of
moduli space of punctured spheres suggests another analogy. Namely, 
in the Matrix Model formulation of 2D Quantum Gravity, it has been 
shown that the theory is described by the Painlev\'e I equation. In 
particular, in \cite{bomama} it has been argued that, while the 
genus expansion corresponds to the weak coupling expansion, the 
expansion labeled by the number of punctures inserted on the sphere 
corresponds to a strong coupling expansion. This similarity, and the 
above remarks, suggest that the expansion at strong coupling 
corresponds to a new kind of moduli space, which is a sort of dual 
of the instanton moduli space. Presumably, this investigation is 
related to the Seiberg-Witten monopole equation \cite{Wittenetal}.

\noindent In conclusion, we have introduced the instanton
universal moduli space for $N=2$ SYM theory with gauge group
$SU(2)$. In particular, we used the recursive structure of the
compactification of instanton moduli space introduced in \cite{M1}
to build, by an inductive limit, an infinite-dimensional moduli
space which encompasses all instanton moduli spaces.

\vspace{.5cm}

\noindent {\bf Acknowledgements.} We would like to thank D.
Bellisai and J. Isidro for many fruitful discussions. Work partly
supported by the European Commission TMR programme
ERBFMRX--CT96--0045.

\end{document}